\documentclass[12pt]{article}

\usepackage{amsmath,amssymb}

\newtheorem{lemma}{Lemma}
\newtheorem{remark}{Remark}
\newtheorem{definition}{Definition}

\newcommand{\pa}{\partial}
\newcommand{\const}{\mathop{const}}

\begin{document}

\title{Remarks on the formation and decay of multidimensional shock
waves}
\author{V.~G.~Danilov\thanks{Moscow Technical University of
Communications and Informatics,\hfill\break 
danilov@miem.edu.ru\break
This work was supported by the Russian Foundation for Basic Research
under grant no.~05-01-00912.}}
\date{}

\maketitle

\begin{abstract}
In this paper, we present a formula describing the 
formation and decay of shock wave type solutions 
in some special cases.
\end{abstract}

In \cite{1} and \cite{2}, in the quadratic and general cases
of convex nonlinearity, we consider the process of formation of
shock waves for scalar conservation laws in the one-dimensional
case. 
Recall that, in the construction suggested in these papers, 
the key role is played by the function $u_1(x)$ determined by
the implicit equation
\begin{equation}
f'(u_1(x))=-Kx+b,
\end{equation}
where $K>0$ and $b$ are constants, and 
$f(u)$ is the nonlinear (convex) density
of the conservation law.

In the present paper, we generalize this construction 
to the multidimensional case. 

The main point is to generalize Eq.~(1).
Recall that the function $u_1(x)$ in (1) 
describes both the shock wave formation and 
the decay of a nonstable step function 
(a rarefaction wave type solution). 

The problems of formation and decay of step functions 
are closely related to each other: 
the change $t\to-t$ allows one 
to use solutions describing the step function formation  
to construct solutions describing the step function decay, 
and conversely.

This procedure is described in detail
(in the framework of the technique used there)
in~\cite{1} in the scalar quadratic case. 

We consider the equation
\begin{equation}
\frac{\pa u}{\pa t}+\sum^{n}_{i=1}\frac{\pa }{\pa x_i}f_i(u)=0,
\end{equation}
where $f_i(u)$ are smooth functions.

To Eq.~(2) there corresponds the system of 
differential equations (equations of characteristics)
\begin{align}
\dot x_i&=f'_i(u),&\qquad x\big|_{t=0}&=x_0,\\
\dot u&=0,        &\qquad u\big|_{t=0}&=u_0(x_0).
\nonumber
\end{align}

We have the following obvious assertion.

\begin{lemma}
\begin{equation}
J\stackrel{\rm def}{=} \det\bigg|\frac{\pa x}{\pa x_0}\bigg|
=t\sum^{n}_{i=1}\frac{\pa^2 f_i(u_0)}{\pa u^2}
\frac{\pa u_0}{\pa x_{0i}}.
\end{equation}
\end{lemma}

The proof readily follows from the relation
$$
\frac{d^2 J}{dt^2}=0.
$$

A generalization of~(1) is based on the fact that 
the derivative of the left-hand side of~(1) with respect to~$x$ 
is exactly the expression under the sign of sum in~(4)
in the one-dimensional case.

Namely, suppose that two smooth surfaces $\Gamma_1$ and
$\Gamma_2$, $\Gamma_1\cap\Gamma_2=\varnothing$, 
are given in a connected simply connected domain~$\Omega$.

We assume that $\Gamma_i$ are determined 
by the equations $x^i_j=\chi^i_j(s)$, 
$s\in{\cal D}\in R^{n-1}$, $i=1,2$, $j=1,\dots,n$.

We also assume that in~$\Omega$ 
there exists a solution $u_1(x)$ of the problem
\begin{gather}
\sum^{n}_{i=1}\frac{\pa^2 f_i(u_0)}{\pa u^2}(u_1)
\frac{\pa u_1}{\pa x_i}+K=0,\\
u_1\big|_{\Gamma_1}=U=\const,\qquad
u_1\big|_{\Gamma_2}=u^0_0=\const,
\nonumber
\end{gather}
where $K=K(s)>0$ is an unknown function
which we seek together with~$u_1(x)$.
It is clear that problem~(5) is the required generalization of~(1).

\begin{remark}\rm
The condition $K=K(s)$ means that the function $K(s)$
is constant on the characteristics corresponding to 
Eq.~(5).
\end{remark}

\begin{remark}\rm
The solvbility of~(5) means that the vector field $f''(u_0)$   
is not singular, i.e., 
$$
|f''|\not=0,
$$
which is an analog of the convexity condition.
\end{remark}

In the multidimensional case,
the function $u_1(x_1,\dots,x_n)$, 
i.e., the solution of~(5), will play the same role 
as the solution of Eq.~(1) in the one-dimensional case.

Problem~(5) is equivalent to the following one:
\begin{align}
\frac{dX_i}{d\tau}&=f''_i(u_1),&\qquad 
X_i\big|_{\tau=0}&=\chi^1_i(s),\\
\frac{du_1}{d\tau}&=-K,&\qquad u_1\big|_{\tau=0}&=U,\qquad i=1,\dots,n.
\nonumber
\end{align}
We have
$$
u_i=U-K(s)\tau.
$$
Let $\tau_0(s)$ be such that 
$X(\tau_0(s),s)\in\Gamma_2$, then 
$$
K(s)=\frac{U-u^0_0}{\tau_0(s)}.
$$

For given $U$ and $u^0_0$,
the condition that $K(s)$ is positive 
implies restrictions on the direction of motion 
along the trajectories determined by (6), 
and the fact that problem~(5) has a solution means that 
$\Gamma_1$ and $\Gamma_2$ are sections of the bundle 
determined by the trajectories of~(6).

To be definite, we assume that $U>u^0_0$. 
Then for $K>0$ the motion along the trajectories of~(6) 
must occur from $\Gamma_1$ to $\Gamma_2$ with increasing~$\tau$.
Otherwise, problem~(5) does not have solutions.

Next, by $\Omega^-$ we denote the domain lying
``before''~$\Gamma_1$, i.e., the domain entered by the
trajectories of system~(6) for $\tau<0$.
By $\Omega^+$ we denote the domain lying
``after''~$\Gamma_2$, i.e., the domain entered by the
trajectories of system~(6) for $\tau>\tau_0(s)$.

By $H^\pm$ we denote the characteristic functions of the
domains~$\Omega^\pm$, 
and by $H^0$ we denote the characteristic function of the domain
$\Omega\setminus\Omega^+\setminus\Omega^-$.

We consider the equation 
\begin{equation}
\frac{\pa u}{\pa t}+\sum\frac{\pa f_i(u)}{\pa x_i}=0
\end{equation}
and set 
\begin{equation}
u\big|_{t=0}=UH^- + u^0_0 H^+ + u_1(x) H^0.
\end{equation}

Next, we must define the concrete geometry of the problem.
For example, we can assume that $\Gamma_1$ and $\Gamma_2$ 
are closed and $\Gamma_2$ is located in the interior of
$\Gamma_1$, or conversely.
However, we shall not do this, but simply assume that we are
interested in the solution of problem (7), (8) in the domain
where it can be obtained from the initial condition 
by using the characteristics.

Clearly, it follows from Lemma~1 and the choice of the function
$u_1(x)$ in (8) 
that the wave turns over on the trajectories 
of the characteristic system corresponding to~(7).

More precisely, 
the trajectories of the characteristic system 
corresponding to~(7),
\begin{equation}
\frac{dx_i}{dt}=f'_i(u_1),\qquad 
x_i\big|_{t=0}=x_0,\qquad i=1,\dots,n,
\end{equation}
such that the point $(x_{10},\dots,x_{n0})=X(s,\tau)$ 
belongs to the trajectory of system~(6) for some fixed~$s$ 
and $0\leq\tau\leq\tau_0(s)$ 
intersect at $t=t_0(s)=1/K(s)$ 
at the point $x=x^*(s)$, $x^*(s)=x(t_0(s),X(s,\tau))$.

Thus, for $t>\max_s t_0(s)$, 
the evolution of the initial condition~(8)
gives a shock type solution of the form 
\begin{equation}
u=U+H(S(x,t))(u^0_0-U),
\end{equation}
where $S\in C^\infty$, $H(s)$ is the Heaviside function,
and the set $S(x,t)=0$ is the shock wave front.

Moreover, at the point $\bar x$ at which the jump occurs,
we have the inequality 
\begin{equation}
u_+-u_-<0,
\end{equation}
where  $u_+$ is the limit value of the solution 
calculated along the trajectory of system~(6) as $x\to\bar x$;
here $x$ corresponds to the value $\tau>\bar\tau$, \
$\bar\tau$ corresponds to $\bar x$, 
and $u_-$ is determined similarly.

Inequality~(11) can be treated as the stability condition for the
jump of the solution to Eq.~(7) in the multidimensional case.

Of course, we here must take into account the above assumption 
on the direction of motion along the trajectories of system~(7).

It is easy to see that the limit $u_+$ can also be calculated 
along the vector $f''_u(u^0_0)$, 
and the limit $u_-$ along the vector $f''_u(U)$.

We agree to denote the limit of $g(x,t)$ as $x\to\bar x$ 
for fixed~$t$ along the vector $X$ by 
$$
(X) \stackrel{\longrightarrow}{\lim_{x\to\bar x}} g(x,t),
$$
and the limit of $g(x,t)$ as $x\to\bar x$ 
along the vector $X$ but in the opposite direction
by 
$$
(X) \stackrel{\longleftarrow}{\lim_{x\to\bar x}} g(x,t).
$$

Then the stability condition for the jump~(11) 
can be written as
\begin{equation}
(f''_{uu}(u^0_0)) \stackrel{\longleftarrow}{\lim_{x\to\bar x}} u(x,t)
-(f''_{uu}(U)) \stackrel{\longrightarrow}{\lim_{x\to\bar x}} u(x,t) 
< 0.
\end{equation}

\begin{definition}\rm
A piecewise constant solution of Eq.~(7) of the form~(10)
is said to be {\it absolutely nonstable\/} 
if the following inequality holds 
at all points $\bar zx\in\{s(\bar x,t)=0\}$ for a fixed~$t$: 
\begin{equation}
(f''_{uu}(u^0_0)) \lim_{x\to\bar x} u(x,t)
-(f''_{uu}(U)) \lim_{x\to\bar x} u(x,t) > 0.
\end{equation}
\end{definition}

It follows from the above that an absolutely nonstable jump
must turn into a solution of the form~(8). 
This construction is completely similar to the one-dimensional
case. More precisely, in the one-dimensional case,
this is described for the case of quadratic nonlinearity 
in~\cite{1}. 
The same also holds for the case of general convex
nonlinearity in the one-dimensional case~\cite{2}.
In the multidimensional case, in fact, the above assumptions
reduce the problem to the one-dimensional problem 
along the trajectories of system~(6).

We note that system (6) can be easily integrated:
\begin{align}
u_1&=U-K(s)\tau,\\
X_i(\tau,s)&=\chi^1_i(s)+\frac1K (f'_i(U)-f'_i(U-K\tau)),
\qquad i=1,\dots,n.
\nonumber
\end{align}
Now let $X^1_0=\chi^1(s_0)$ be an arbitrary point on
$\Gamma_1$. By $X^2_0=X(\tau_0(s_0),s_0)$ we denote the point of
intersection of the trajectory of system (6) with $\Gamma_2$.
Let $x(X^i_0,t)$, $i=1,2$, be solutions of system (9) 
such that 
$$
x\big|_{t=0}=X^i_0.
$$
Then, by (14), we have
\begin{align}
x(X^1_0,t)-x(X^2_0,t)
&=X^1_0-X^2_0+t(f'(U)-f'(u^0_0))\\
&=\frac1K[f'(U)-f'(u^0_0)](Kt-1).
\nonumber
\end{align}

In the construction of multidimensional nonlinear waves, 
an important role is played by the level surface $\Gamma^t$
of the solution~\cite{1}.
These surfaces are determine by the relations
$\Gamma^t=\{t=\psi(x)\}$, where $\psi(x)$ is the desired unknown
function whose zero-level surface is assumed to be given
(in our problem, these are the surfaces $\Gamma_1$ 
and $\Gamma_2$).

Clearly, if, for example, $\Gamma^0=\Gamma_1$, 
then $\Gamma^t_1$ is the set of the endpoints 
of the trajectories of system~(9) 
starting on $\Gamma_1$ at time~$t$. 
In this case, the function $\psi_1(x)$ is the time 
required for the trajectory starting 
at a point $X_{10}\in\Gamma_1$ to come to the point~$X$. 
Similarly, we determine $\psi_2(x)$ and $\Gamma^t_2$.

Let us consider the expression
$$
[\psi_1(x)-\psi_2(x)]\bigg|_{\Gamma^t_1}
\equiv t-\psi_2(x)\bigg|_{\Gamma^1_t},
$$
where the restriction means that $x$ is a point on $\Gamma^1_t$.
We have
$$
x=X_{10}+\psi_1 f'(U)=X_{20}+\psi_2 f'(u^0_0),
$$
where $X_{10}\in\Gamma_1$ and $X_{20}\in\Gamma_2$ 
are some initial point of the trajectories~(9). 

We have
$$
X_{10}-X_{20}=\psi_2 f'(u^0_0)-\psi_1 d'(U).
$$
Next, we have $X_{10}=\chi^1(s)$ for some~$s$.
We denote $X_{20}=X_1(\tau_0(s),s)$. 
By~(15), we have
$$
(\psi_2-\psi_1)\bigg|_{\Gamma^1_t}f'(U)
=X_{02}-X_{20}+\frac1K[f'(U)-f'(u^0_0)](Kt-1).
$$
Now let $t=1/K$ and $x\in \Gamma^t_1\cap \Gamma^t_2$. 
Then $X_{20}=X_{02}$.
In general, we have
$$
(\psi_2-\psi_1)\bigg|_{\Gamma_t}
=\frac{(Kt-1)K^{-1}\langle f'(U),f'(U)-f'(u^0_0)\rangle}
{\|f'(U)\|^2},
$$
where $\langle,\rangle$ is the inner product in $\mathbb{R}^n$. 
Similarly, we define the quantity
$$
(\psi_2-\psi_1)\bigg|_{\Gamma_{1t}}.
$$

Everything said above is an analog of the Introduction in~\cite{2}.
In the present text, we restrict ourselves to this and only note
that we have prepared everything necessary to construct 
the multidimensional analog of the weak asymptotic solution
given in~\cite{2}, which describes the formation of a shock wave. 
We shall present this in detail in the next paper. 

In conclusion, we formulate the solution of the problem
concerning the decay of a nonstable step-function.

Suppose that the initial data for Eq.~2 have the form~(10)
for $t=0$, and the vector fields $f'(U)$, $f'(u^0_0)$ and
$f''(U)$, $f''(u^0_0)$ 
are transversal to the surface $\Gamma_0=\{S(x,0)=0\}$.
Next,   
we assume that inequality (13) holds at the points of
$\Gamma_0$. 
Then there exists a $\bar t>0$ such that for 
$t\in[0,\bar t]$, the solution of Eq.~(2) with the initial
condition 
$$
u\big|_{t=0}=U+H(S(x,0))(u^0_0-U)
$$
has the form~(8).


\begin{thebibliography}{99}

\bibitem{1} 
V.~Danilov,
{\it Generalized solutions describing singularity interaction},
IJMMS, 29:8 (2002), 481--494.

\bibitem{2}
V.~Danilov and D.~Mitrovich,
{\it Weak asymptotics of shock wave formation process},
Journal of Nonlinear Analysis, Theory, Methods, and
Applications, 61 (2005), 613--635.

\end{thebibliography}
\end{document}